\documentclass[twocolumn,floatfix,eqsecnum,showpacs]{revtex4}
\usepackage{graphicx}
\usepackage{amssymb}
\usepackage{amsmath}
\usepackage{bm}

\pdfoutput=1

\newcommand{\aaa}{\mathcal{A}}
\newcommand{\jj}{J}
\newcommand{\MM}{\mathcal{M}}

\begin{document}

\title{Pseudodiffusive transmission of nodal Dirac fermions through a clean \textit{d}-wave superconductor}
\author{J. K. Asb\'{o}th, A. R. Akhmerov, A. C. Berceanu, and C. W. J. Beenakker}
\affiliation{Instituut-Lorentz, Universiteit Leiden, P.O. Box 9506, 2300 RA Leiden, The Netherlands}
\date{September 2009}
\begin{abstract}
We calculate the transmission of electrons and holes between two normal-metal electrodes (N), separated over a distance $L$ by an impurity-free superconductor (S) with \textit{d}-wave symmetry of the order parameter. Nodal lines of vanishing excitation gap form ballistic conduction channels for coupled electron-hole excitations, described by an anisotropic two-dimensional Dirac equation. We find that the transmitted electrical and thermal currents, at zero energy,  both have the pseudodiffusive $1/L$ scaling characteristic of massless Dirac fermions --- regardless of the presence of tunnel barriers at the NS interfaces. Tunnel barriers reduce the slope of the $1/L$ scaling in the case of the electrical current, while leaving the thermal current unaffected.
\end{abstract}
\maketitle

\section{Introduction}
\label{intro}

Pseudodiffusive transmission refers to the $1/L$ scaling of the electrical current transmitted over a distance $L$ through a clean sheet of undoped graphene \cite{Bee08}. The same $1/L$ scaling characterizes diffusion in a random potential, but now it applies in the absence of any disorder. There is a large number of theoretical \cite{Kat06,Two06,Akh07,Pra07,Sch07,Bla07,Cse07,Cre07,Tit07,Mog09,Die09} and experimental \cite{Mia07,DiC08,Dan08} studies of this phenomenon, which is understood as a general property of massless Dirac fermions in the limit of vanishing excitation energy. The optical analogue in a photonic crystal with a Dirac spectrum has been studied as well \cite{Sep07,Sep08,Zha08a,Zha08b}.

Layered superconductors with a \textit{d}-wave symmetry of the order parameter (notably the high-$T_{c}$ cuprates \cite{Har95}) form an altogether different system in which massless Dirac fermions are known to exist \cite{Lee93,Dur00,Alt02}. These are so-called nodal fermions, located in the two-dimensional Brillouin zone near the intersections (nodal points) of the Fermi surface with lines (nodal lines) of vanishing excitation gap. Elastic mean free paths $l$ as large as $4\,\mu{\rm m}$ have been reached in ${\rm YBa}_{2}{\rm Cu}_{3}{\rm O}_{7-\epsilon}$ single-crystals \cite{Har06}, much larger than the superconducting coherence length $\xi_0\simeq 2\,{\rm nm}$. It is the purpose of this work to demonstrate theoretically the pseudodiffusive $1/L$ scaling of the transmission through a \textit{d}-wave superconductor over the range of lengths between $\xi_0$ and $l$. This anomalous scaling was not noticed in earlier studies of similar systems \cite{Don04,Tak06,Her09}.

The problem is interesting from a conceptual point of view, because it highlights both the differences and similarities between Dirac fermions produced by a bandstructure (as in graphene or photonic crystals) or produced by a \textit{d}-wave order parameter. In undoped graphene, the transmitted electrical current $I$ in response to a voltage difference $V$ scales as \cite{Kat06,Two06}
\begin{equation}
I=\frac{4e^{2}}{h}V\frac{W}{\pi L}.\label{Igraphene}
\end{equation} 
The length $L$ over which the current is transmitted should be large compared to the Fermi wave length $\lambda_{F}$ in the metal contacts, but small compared to the mean free path $l$. The length $L$ should also be small compared to the transverse width $W$ of the graphene sheet (to avoid edge effects). Potential barriers (smooth on the scale of the lattice constant) at the interfaces between the metal contacts and the graphene sheet have no effect on the current, because of the phenomenon of Klein tunneling \cite{Tit07}. 

For the \textit{d}-wave superconductor, we find a transmitted electrical current per layer equal to
\begin{equation}
I=\frac{2e^{2}}{h}V\frac{W}{\pi L}\frac{v_{F}^{2}+v_{\Delta}^{2}}{v_{F}v_{\Delta}}\frac{\Gamma_1}{(2-\Gamma_1)}\frac{\Gamma_2}{(2-\Gamma_2)},\label{Idwave}
\end{equation}
for $\xi_0\ll L\ll l,W$. Here $\Gamma_{1,2}\in(0,1)$ are the tunnel probabilities through the potential barriers at the two  normal-metal--superconductor (NS) interfaces. The Dirac equation for nodal fermions is anisotropic \cite{Lee93}, with different velocities $v_{F}$ and $v_{\Delta}$ parallel and perpendicular to the nodal lines. This anisotropy (with $v_{F}/v_{\Delta}\approx 15$ in ${\rm YBa}_{2}{\rm Cu}_{3}{\rm O}_{7-\epsilon}$) increases the slope of the $1/L$ scaling. Remarkably enough, the anisotropy does not introduce a dependence of the transmitted current on the angle $\alpha$ between the direction of the current and the nodal lines. The result \eqref{Idwave} holds generically for any orientation, except for a narrow range of angles of order $\xi_0/L$ around $\alpha=0\;({\rm mod}\;\pi/4)$.

The tunnel barriers reduce the slope of the $1/L$ scaling of the transmitted electrical current \eqref{Idwave}, by a factor $\Gamma_1 \Gamma_2/4$ for small tunnel probabilities. This does not imply that the nodal fermions are only weakly transmitted, but rather that the transmission probabilities for transmission as an an electron or as a hole are almost the same for $\Gamma_{1,2}\ll 1$. Indeed, we find that the electrical shot noise power $P$ as well as the transmitted thermal current $I_{\rm thermal}$ (both of which do not depend on the sign of the carriers charge) remain finite in the limit $\Gamma_{1,2}\rightarrow 0$. We interpret this result in terms of a resonant coupling via the nodal lines of the mid-gap states \cite{Hu94,Kas00} extended along the two NS interfaces. We also find, quite surprisingly, that the thermal conductivity is \emph{independent} of the tunnel probabilities $\Gamma_{1,2}$.

The outline of this paper is as follows. In Sec.\ \ref{nodalT} we formulate the scattering problem and calculate the transfer matrix of the nodal Dirac fermions through the \textit{d}-wave superconductor. The matching of wave functions at the interface with the metal electrodes is done in Sec.\ \ref{wavematching}, both for ideal NS interfaces and for interfaces containing a tunnel barrier. The transmission matrix of electrons and holes follows in Sec.\ \ref{transmission}. We then apply this result to the calculation of transport properties: the electrical current (Sec.\ \ref{electric}), the thermal current (Sec.\ \ref{thermal}), and the electrical shot noise (Sec.\ \ref{shotnoise}). We conclude in Sec.\ \ref{discuss} with a discussion of our results and an outlook.

\section{Transfer matrix for nodal fermions}
\label{nodalT}

\subsection{Anisotropic Dirac equation}
\label{anisoDirac}

\begin{figure}[tb]
\centerline{\includegraphics[width=0.9\linewidth]{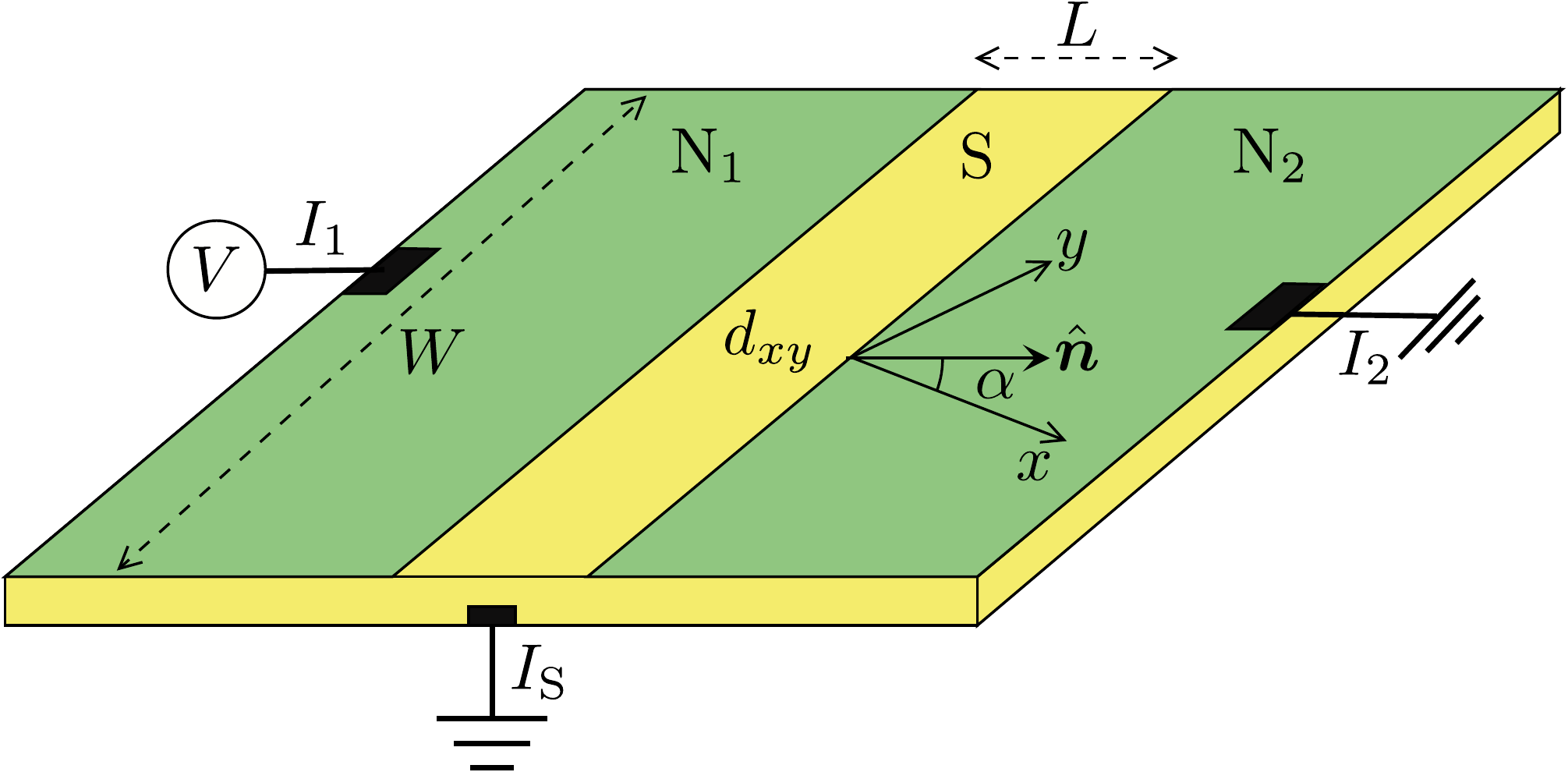}}
\caption{\label{fig_layout}
Geometry to measure the transmission of nodal fermions through a $d$-wave superconductor. A current $I_{1}$ is injected into the superconductor from metal contact ${\rm N}_{1}$ (at a voltage $V$) and drained to ground via the superconductor (current $I_{S}$) or via a second metal contact ${\rm N}_{2}$ (current $I_{2}$). If the separation $L$ of the metal contacts is large compared to the superconducting coherence length $\xi_0$, the current $I_{2}$ is predominantly due to transmission parallel to the nodal lines $x=0$ or $y=0$ of vanishing excitation gap .
}
\end{figure}

We consider a two-dimensional spin-singlet superconductor (S), connecting two normal metal contacts with parallel NS interfaces, separated by a distance $L$. 
The transverse dimension $W$ of the superconducting strip (in the $x$-$y$ plane) is assumed to be large compared to $L$, in order to avoid edge effects.  
The order parameter $\Delta(\bm{k})$ is assumed to have $d_{xy}$ symmetry: it vanishes for wave vectors along two nodal lines, which are taken to be the $x$ and the $y$ axis. 
All our results also apply to $d_{x^2-y^2}$-superconductors, for our purposes, a simple $\pi/4$ rotation relates the two systems.   
To be specific, the $x$-$y$ plane can represent a single ${\rm CuO}_{2}$ layer of a cuprate superconductor \cite{Har95}, with the $[100]$ direction at an angle $\pi/4$.  

\begin{figure}[tb]
\centerline{\includegraphics[width=0.7\linewidth]{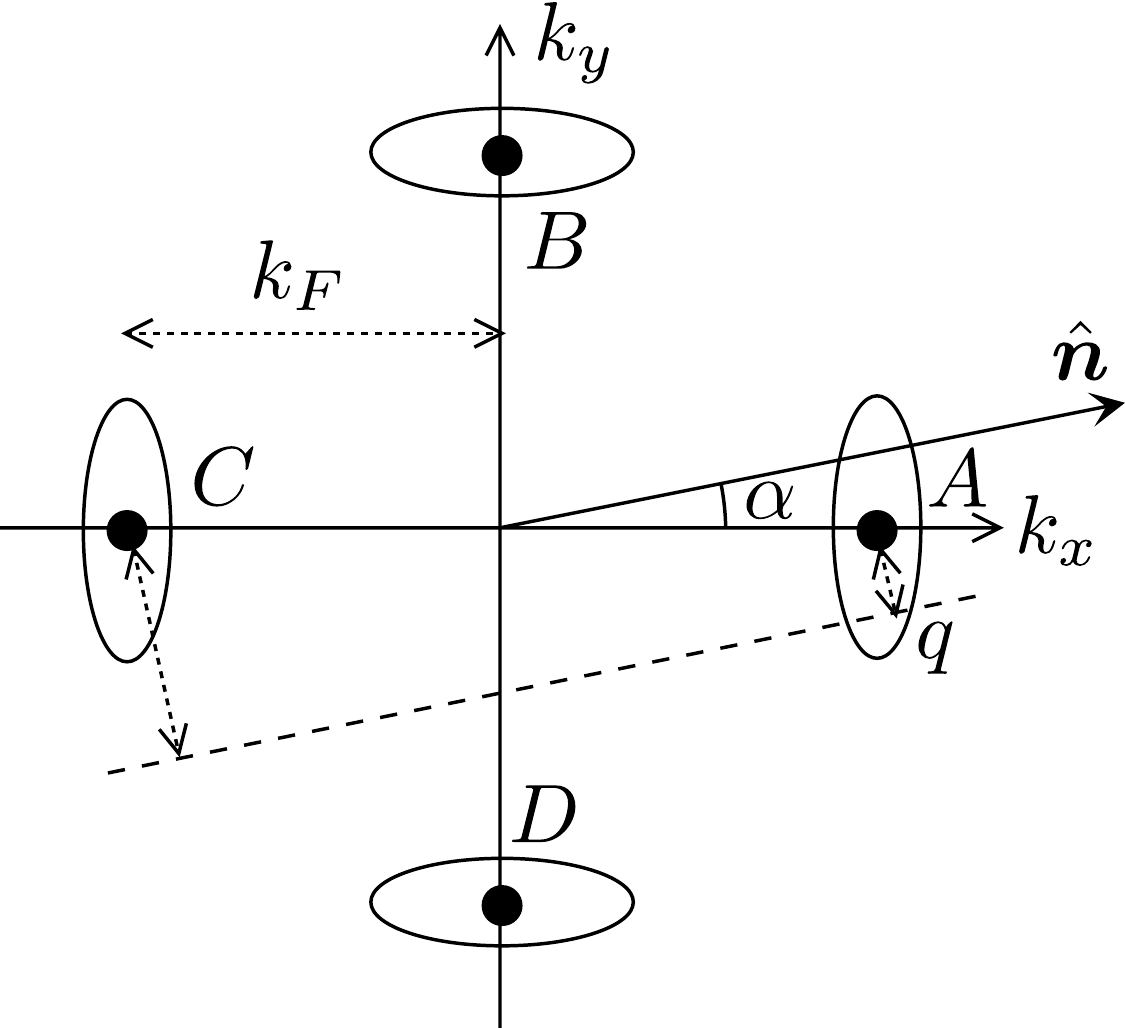}}
\caption{\label{fig_Brillouin}
Ellipsoidal equal-energy contours of low-energy excitations in the Brillouin zone of a superconductor with $d_{xy}$ symmetry. Long and short axes have ratio $v_{F}/v_{\Delta}$. The contours are centered at the four nodal points (solid dots), where the order parameter vanishes on the Fermi surface. The normal $\hat{\bm{n}}$ to the NS interfaces is indicated. The dashed line, displaced from the nearest nodal point by $q$, indicates points of constant wave vector component parallel to the interface.
}
\end{figure}

Low-energy excitations in the superconductor are found in the Brillouin zone near the four intersections $(\pm k_{F},0)$, $(0,\pm k_{F}$) of the Fermi surface with the nodal lines of the order parameter. (These nodal points are labeled $A,B,C,D$ in Fig.\ \ref{fig_Brillouin}.) Around these points, both the pair potential $\Delta(\bm{k})$ and the kinetic energy can be linearized: the dynamics of the nodal fermions is governed by an anisotropic Dirac equation \cite{Lee93,Dur00,Alt02}. For example, near node $A$ at $(k_{F},0)$ this can be written in the form
\begin{equation}
\begin{pmatrix}
-i v_{F}\partial_{x}&-i v_{\Delta}\partial_{y}\\
-i v_{\Delta}\partial_{y}&i v_{F}\partial_{x}
\end{pmatrix}
\begin{pmatrix}
\Psi_{e}\\
\Psi_{h}
\end{pmatrix}
=\varepsilon
\begin{pmatrix}
\Psi_{e}\\
\Psi_{h}
\end{pmatrix},
\label{Dirac1}
\end{equation}
or more compactly with the help of Pauli matrices,
\begin{equation}
-i[v_{F}\sigma_{z}\partial_{x}+ v_{\Delta}\sigma_{x}\partial_{y}]\Psi=\varepsilon\Psi.
\label{Dirac2}
\end{equation}
We have set $\hbar$ to unity, restoring units in the final expressions.  
The spinor $\Psi=(\Psi_{e},\Psi_{h})$ contains the envelope wave functions of electron and hole excitations (slowly varying on the scale of the Fermi wavelength $\lambda_{F}=2\pi/k_{F}$). The Fermi velocity $v_{F}$ is larger than the velocity $v_{\Delta}=\Delta_{0}/\hbar k_{F}$ by a factor of order $\xi_0/\lambda_{F}$ (with $\xi_0=\hbar v_{F}/\Delta_{0}$ the superconducting coherence length), which is in the range 10--20 for cuprate superconductors. The equal-energy contours in the Brillouin zone of the nodal fermions thus have an elongated ellipsoidal shape,
\begin{equation}
\varepsilon(\delta\bm{k})=\sqrt{( v_{F}\delta k_{x})^{2}+( v_{\Delta}\delta k_{y})^{2}},\label{epsk}
\end{equation}
as a function of the displacement $\delta\bm{k}$ of the wave vector from the nodal point.

\subsection{Transfer matrix}
\label{tmatrix}

Since the system is translation invariant along the NS interfaces, the component of the wave vector along these interfaces, $q=-\delta k_{x}\sin\alpha+\delta k_{y}\cos\alpha$, is a conserved quantity. Here $\alpha$ is the angle between the normal to the NS interface and the nodal line pointing to node $A$, which we restrict to  $-\pi/4\le \alpha \le \pi/4$ without loss of generality.
Moreover, since mirror reflection along the NS interface, followed by the transformation $\Delta(\bm{k})\rightarrow -\Delta(\bm{k})$, while leaving all the other parameters unchanged, maps $\alpha$ on $-\alpha$, we can further restrict $\alpha$ to $0\le\alpha\le \pi/4$. In all our formulas, to obtain the corresponding formulas for $-\alpha$, replace $q$ by $-q$ and $v_\Delta$ by $-v_\Delta$.   

We write $\Psi(\bm{r})=\Psi(s)e^{iqs'}$, with $s\in(0,L)$ the coordinate perpendicular to the NS interfaces and $s'$ the coordinate parallel to them. We substitute $\Psi(\bm{r})$ into Eq.~\eqref{Dirac1} and find that the spinor $\Psi(s)$ satisfies the wave equation
\begin{equation}
\left[-i \jj \partial_{s}+ 
\jj_1
q \right] \Psi(s)=\varepsilon\Psi(s),
\label{Dirac3}
\end{equation}
where $\partial_{s}$ is differentiation perpendicular to the NS interface, and
$\jj$ and $\jj_1$ are the operators of particle current perpendicular and parallel to the NS interface, 
\begin{align}
\jj &= v_{F} \sigma_z \cos\alpha + v_{\Delta}\sigma_x\sin\alpha\, ,
\label{j_def}\\
\jj_1 &= v_\Delta \sigma_x \cos \alpha  - v_F \sigma_z \sin \alpha.
\end{align}
We note that the operator $\jj$ squares to a scalar, its magnitude giving the particle velocity $v_\alpha$ perpendicular to the NS interface:
\begin{equation}
v_{\alpha}^2 =\jj^2 = v_{F}^{2}\cos^{2}\alpha+v_{\Delta}^{2}\sin^{2}\alpha.\label{vbardef}
\end{equation}

To solve Eq.~\eqref{Dirac3}, we multiply it by $\jj/v_\alpha^2$ and rearrange to obtain 
\begin{equation}
\partial_s \Psi(s) = i \aaa_0 \Psi(s),
\label{diff_A}
\end{equation}  
with 
\begin{equation}
\aaa_0 =  
q\, \frac{\sin 2\alpha}{2}\, \frac{v_F^2-v_\Delta^2}{v_\alpha^2} + \frac{\varepsilon }{v_\alpha^2} \jj - i q\, 
\frac{v_F v_\Delta}{v_\alpha^2}\, \sigma_y .
\label{def_A}
\end{equation} 
The solution to Eq.~\eqref{Dirac3} can then be written as
\begin{align}  
\Psi(s_0+s) &= \MM_s \Psi(s_0); \quad 
&\MM_s &= \exp [i \aaa_0 s] ,
\label{MSdef}
\end{align}  
where the second equation defines the \emph{transfer matrix} $\MM_{s}$.
As expected, the particle current $\jj$ is conserved by Eq.~\eqref{Dirac3}: $\partial_s \left[\Psi^\dagger(s) \jj\Psi(s)\right] = 
\Psi^\dagger(s) [-i\aaa_0^\dagger \jj +i \jj \aaa_0] \Psi(s)=0$. 

\section{Wave matching at the NS interfaces}
\label{wavematching}

At the two NS interfaces the coupled electron-hole excitations in the superconductor are converted into uncoupled electrons and holes in the normal metal. We thus need to match, at $s=0$ and $s=L$, the envelope wave functions $\Psi=(\Psi_{e},\Psi_{h})$ of the nodal fermions in S to the Bloch wave functions $\Phi=(\Phi_{e},\Phi_{h})$ of free fermions in N.
This is similar to the matching of Dirac equation to Helmholtz equation considered in the context of transmission through a photonic crystal \cite{Sep07}.
Translational invariance parallel to the NS interfaces requires that the coupling conserve the wave vector component $q$ parallel to the interfaces. 
Particle flux conservation imposes further constraints, as we determine here.

\subsection{Particle flux conservation at the NS interface}
\label{fluxcons}

At the surface of the superconductor, the order parameter $\Delta$ attains its bulk value over a short length scale, the healing length $l_0$.
The two-component wave function on the S side of the interface (at $s=l_0$) can be linked to that on the N side ($s=0$) by an \emph{interface matrix} $\MM_{NS}$, defined by
\begin{equation} 
\Psi(l_0) = 
\MM_{NS} \Psi(0).
\label{MNSdef}
\end{equation}
In the normal metal, the operator of particle flux perpendicular to the NS interface can be written as 
\begin{align}  
j_N = v_N \sigma_z,
\end{align}  
with $v_N$ possibly different from $ v_F \cos \alpha$ because of a Fermi energy mismatch. 
The requirement of particle flux conservation reads
\begin{equation}
\Psi(0)^\dagger  v_N \sigma_z  \Psi(0)  = 
\Psi(0)^\dagger  \MM_{NS}^{\dagger} \jj \MM_{NS}  \Psi(0).
\label{jNdef}
\end{equation} 

To derive the most general form of the interface matrix fulfilling this requirement, notice that a unitary rotation 
through angle $\theta$, where 
\begin{align}
\theta = \arctan \left[ \frac{v_\Delta}{v_F} \tan \alpha  \right],
\label{theta_def}
\end{align}
 transforms $\jj$ into $\sigma_z$ up to a scalar factor: 
\begin{equation}
v_\alpha \sigma_z = \exp[i\theta \sigma_y /2]  \jj \exp[-i\theta \sigma_y /2]. 
\label{jNdef}
\end{equation}
This allows us to write the interface matrix as 
\begin{equation} 
\MM_{NS}  = \sqrt{\frac{v_N}{ v_\alpha}} \exp[-i\theta \sigma_y /2] 
\MM_0,
\label{MNSdef}
\end{equation} 
where $\MM_0$ is a $2\times 2$ matrix fulfilling a generalized unitarity condition, 
\begin{equation}
M_{0}^{-1}=\sigma_{z}M_{0}^{\dagger}\sigma_{z}.\label{symplectic}
\end{equation}

Eq.\ \eqref{symplectic} restricts $M_{0}$ to a three-parameter form
\begin{equation}
M_{0}=e^{\beta_x\sigma_{x}}e^{\beta_y\sigma_{y}}e^{i\beta_z\sigma_{z}}
\label{M0}
\end{equation}
(ignoring an irrelevant scalar phase factor), with arbitrary real parameters $\beta_x,\beta_y,\beta_z$.
To understand better where the nontrivial interface matrix arises from, and to show that we may set  
$\beta_x=\beta_y=\beta_z=0$, we have to extend the Dirac equation \eqref{Dirac3} to the interface layer, where 
$v_\Delta$ varies in space. This is done in Appendix \ref{MNS_derivation}.

So far we have considered only \textit{intranode} scattering at the NS interface. We refer to such an interface as an ``ideal interface''. A nonideal interface contains a tunnel barrier, which introduces \textit{internode} scattering. We will consider the transfer matrices through the \textit{d}-wave superconductor for both cases in the next two subsections.

\subsection{Transfer matrix with ideal NS interfaces}
\label{tmideal}

The complete transfer matrix for a strip of {\textit d}-wave superconductor with ideal NS interfaces reads  
\begin{align}
{\MM}_{\rm ideal}&= \MM_{NS}^{-1} \MM_L \MM_{NS},
\label{MM_ideal_def}
\end{align}
where $\MM_L$ is the $\MM_s$ from Eq.~\eqref{MSdef} with $s=L$, describing propagation inside the superconductor, and $\MM_{NS}$ from Eq.~\eqref{MNSdef}, with $\MM_0=1$, describes an NS interface. 
Upon substitution, we obtain  
\begin{align}
\MM_{\rm ideal} &= e^{i\phi_{\alpha}(q)}\exp\biggl(\frac{iL\varepsilon}{v_{\alpha}} \sigma_{z} 
+ qL \,\frac{v_{F}v_{\Delta}}{v_{\alpha}^{2}} \sigma_{y}\biggr)\nonumber\\
&=e^{i\phi_{\alpha}(q)}\biggl[\cosh[\kappa_{\alpha}(q) L]\nonumber\\
&\quad\mbox{}+\frac{\sinh[\kappa_{\alpha}(q)L]}{v_{\alpha}^{2}\kappa_{\alpha}(q)}\left(i\varepsilon v_{\alpha}\sigma_{z} + qv_{F} v_{\Delta}\sigma_{y}\right)\biggr],
\label{Mideal}
\end{align}
with the definitions
\begin{align}
&\kappa_{\alpha}(q)=\sqrt{(qv_{F}v_{\Delta}/v_{\alpha}^{2})^{2}-(\varepsilon/v_{\alpha})^{2}},\label{kappadef}\\
&\phi_{\alpha}(q)= qL \frac{v_{F}^{2}-v_{\Delta}^{2}}{v_\alpha^2}\frac{\sin 2\alpha}{2}.\label{phialphadef}
\end{align}

Notice, how -- as a result of accounting for the two 
NS interfaces -- the transfer matrix has simplified from that of Eq.~\eqref{MSdef}. 
The change is that the particle flux operator $\jj$ in Eq.~\eqref{def_A} is replaced by 
$v_\alpha \sigma_z$ in Eq.~\eqref{Mideal}. 
Also note that the determinant of the transfer matrix has norm one, $\mathrm{Det} \MM_{\rm ideal} = e^{2i\phi_\alpha (q)} $, 
as required by the generalized unitarity relation
\begin{equation}
\MM^{-1}=\sigma_z\MM^\dagger \sigma_z,
\label{genunitary}
\end{equation}
which holds for any transfer matrix as a consequence of particle current conservation.

To appreciate the effects of the Dirac cone anisotropy, we can perform a linear transformation on our system to obtain one with an isotropic Dirac cone:  
contraction along the nodal line by a factor $v_\Delta/v_\alpha$, and expansion perpendicular to it by a factor $v_F/v_\alpha$. 
The dispersion of the new, isotropic Dirac cone has a single velocity parameter $v = v_F v_\Delta / v_\alpha$. 
The superconducting strip is deformed by the transformation: its width $W$ is unchanged, but its length $L$ becomes 
\begin{align} 
L_\alpha = L \frac{v_F v_\Delta}{v_\alpha^2}, 
\label{Lalpha_def}
\end{align}
an effective propagation length we define here for later use. 


The matrix \eqref{Mideal} derived above is the transfer matrix for nodal fermions near point $\bm{k}_A=(k_{F},0,0)$ on the Fermi surface, with $q\equiv q_A=(\bm{k}-\bm{k}_A)\cdot(\bm{\hat{z}}\times\bm{\hat{n}})$ the transverse wave vector component relative to $k_A$. Similarly, the transfer matrices near each of the four nodal points can be written as
\begin{subequations}
\label{MABCD}
\begin{align}
&\MM_A=e^{i\phi_{\alpha}(q_A)}\exp\biggl(\frac{iL\varepsilon}
{v_{\alpha}}\sigma_{z} + \frac{q_ALv_{F}v_{\Delta}}{v_{\alpha}^{2}}\sigma_{y}\biggr),
\label{MA}\\
&\MM_{B}=e^{-i\phi_{\pi/2-\alpha}(q_{B})}\exp\biggl(\frac{iL\varepsilon}
{v_{\pi/2-\alpha}}\sigma_{z} - \frac{q_{B}Lv_{F}v_{\Delta}}{v_{\pi/2-\alpha}^{2}}\sigma_{y}\biggr),
\label{MB}\\
&\MM_{C}=e^{i\phi_{\alpha}(q_{C})}\exp\biggl(-\frac{iL\varepsilon}
{v_{\alpha}}\sigma_{z} + \frac{q_{C}Lv_{F}v_{\Delta}}{v_{\alpha}^{2}}\sigma_{y}\biggr),
\label{MC}\\
&\MM_{D}=e^{-i\phi_{\pi/2-\alpha}(q_{D})}\exp\biggl(-\frac{iL\varepsilon}
{v_{\pi/2-\alpha}}\sigma_{z} - \frac{q_{D}Lv_{F}v_{\Delta}}{v_{\pi/2-\alpha}^{2}}\sigma_{y}\biggr)
\label{MD}.
\end{align}
\end{subequations}
The basis at each nodal point is the same spinor $(\Phi_{e},\Phi_{h})$, but the electron states $\Phi_{e}$ are ``right-movers'' (propagating from ${\rm N}_{1}$ to ${\rm N}_{2}$) at nodal points $A$ or $B$ and ``left-movers'' (from ${\rm N}_{2}$ to ${\rm N}_{1}$) at nodal points $C$ and $D$.

\subsection{Nonideal interfaces}
\label{nonideal}

The complete Fermi surface of the normal metal (N) might differ in many ways from that of the superconductor. However, when we study transport near a specific nodal point, due to transverse momentum conservation, we can effectively reduce the Fermi surface to the two $\bm{k}$ points where transverse momentum has the same value as at the nodal point. 
These two $\bm{k}$ points in N each couple to different nodal points in S, for example to nodal points $A$ and $C$ in Fig. 2. A nonideal NS interface couples different nodal points, by reversing the component of the momentum perpendicular to the interface. Such \textit{internode scattering} may be caused by an insulating layer at the NS interface, or it may result from the Fermi velocity mismatch between N and S. 
Note that only internode scattering is possible in the absence of superconducting order --- any \textit{intranode} scattering has to happen inside the superconductor. 

We will generically describe a nonideal NS interface by a tunnel barrier, with tunnel probability $\Gamma$ (which we take mode independent for simplicity).
For $|\alpha| \lesssim \xi_0/L $, the tunnel barrier couples electrons near nodal points $A$ and $C$. The transfer matrix $\MM_{\Gamma}(s_{0})$ for a tunnel barrier at position $s_{0}$, defined by
\begin{equation}
\begin{pmatrix}
\Phi_{e,A}\\
\Phi_{e,C}
\end{pmatrix}_{s_{0}^{+}}
=\MM_{\Gamma}(s_{0})
\begin{pmatrix}
\Phi_{e,A}\\
\Phi_{e,C}
\end{pmatrix}_{s_{0}^{-}},\label{MACedef}
\end{equation}
has the form
\begin{equation}
\MM_{\Gamma}(s_{0})=\sqrt{\frac{1}{\Gamma}}\begin{pmatrix}
1&e^{-i\phi(s_{0})}\sqrt{1-\Gamma}\\
e^{i\phi(s_{0})}\sqrt{1-\Gamma}&1
\end{pmatrix},\label{MAC}
\end{equation}
with $\phi(s)=2k_{F}s\cos\alpha$.

The tunnel barrier at $s_{0}$ also couples holes near nodal points $A$ and $C$, with transfer matrix
\begin{equation}
\begin{pmatrix}
\Phi_{h,C}\\
\Phi_{h,A}
\end{pmatrix}_{s_{0}^{+}}
=\MM_{\Gamma}^{\ast}(s_{0})
\begin{pmatrix}
\Phi_{h,C}\\
\Phi_{h,A}
\end{pmatrix}_{s_{0}^{-}},\label{MAChdef}
\end{equation}
(The basis states are chosen such that the upper component is a right-mover and the lower component a left-mover.)

Finally, we can write down the full transfer matrix of the superconducting strip, in the basis $(\Phi_{e,A},\Phi_{h,A},\Phi_{e,C},\Phi_{h,C})$, including nonideal contacts with tunneling probabilities $\Gamma_1$ at $s=0$ and $\Gamma_2$ at $s=L$. It is obtained by matrix multiplication,
\begin{align}
\MM={}&V\begin{pmatrix}
\MM_{\Gamma_2}(L)&0\\
0&\MM_{\Gamma_2}^{\ast}(L)
\end{pmatrix}V^{\dagger}\begin{pmatrix}
\MM_A&0\\
0&\MM_C
\end{pmatrix}\nonumber\\
&\cdot V\begin{pmatrix}
\MM_{\Gamma_1}(0)&0\\
0&\MM_{\Gamma_1}^{\ast}(0)
\end{pmatrix}
V^{\dagger},
\label{Mfull}
\end{align}
with $V$ a unitary matrix that switches bases from $(\Phi_{e,A},\Phi_{e,C},\Phi_{h,C},\Phi_{h,A})$ to $(\Phi_{e,A},\Phi_{h,A},\Phi_{e,C},\Phi_{h,C})$:
\begin{equation}
V=\begin{pmatrix}
1&0&0&0\\
0&0&0&1\\
0&1&0&0\\
0&0&1&0
\end{pmatrix}.\label{Omegadef}
\end{equation}

If $\pi/4-\alpha \lesssim \xi_0/L$, the nodal point $A$ is coupled to the nodal point $D$, 
so the above formulas still hold, with $C$ replaced by $D$.

If both $\alpha\gg \xi_0/L$ and $\pi/4-\alpha \gg \xi_0/L$, 
the tunnel barriers at the interfaces do not couple nodal point $A$ to any other nodal points.
Since we assume $\xi_0/L \ll 1$, this case of misaligned nodes is the generic case.  
In that case, $\MM_C$ is to be replaced by the singular transfer matrix $\MM_{\rm Andreev}$ corresponding to Andreev reflection with reflection amplitude $-i$,
\begin{equation}
\MM_{\rm Andreev} = \lim_{z\to\infty} e^{-z \sigma_y}.
\label{MClimit}
\end{equation}
Since $\MM_{\rm Andreev}$ is also the $q_{C}\rightarrow-\infty$ limit of 
$\MM_C(q_C)$ in Eq.\ \eqref{MC} (up to an irrelevant phase factor), Eq.\ \eqref{Mfull} is valid as it stands for misaligned nodes as well. 

\section{Transmission amplitudes}
\label{transmission}

\subsection{Ideal interfaces}
\label{tideal}

Referring to the geometry of Fig.\ \ref{fig_layout}, a scattering state (for a given value of $q$) has the form  $\Phi(0)=\bigl(1,r_{he}\bigr)$ at the normal side of the left NS interface and $\Phi(L)=\bigl(t_{ee},0\bigr)$ at the normal side of the right NS interface. The complex number $r_{he}$ is the amplitude for Andreev reflection (from electron to hole) and the complex number $t_{ee}$ is the amplitude for electron transmission. 
We calculate this transmission amplitude using the relation  
\begin{equation}
t_{ee}=\left([\MM_A^{-1}]_{11}\right)^{-1} = \left([\MM_A^\dagger]_{11} \,\right)^{-1},
\label{tMrelation}
\end{equation}
where the first equality follows from $\Phi(0) = \MM_A^{-1} \Phi(L)$, and the second equality from particle current conservation,
Eq.~\eqref{genunitary}.

Substitution of Eq.\ \eqref{MA} gives the expression
\begin{equation}
t_{ee}=e^{i\phi_{\alpha}(q_A)}\left[\cosh[\kappa_{\alpha}(q_A) L]-\frac{i\varepsilon\sinh[\kappa_{\alpha}(q_A) L]}{v_{\alpha}\kappa_{\alpha}(q_A)}\right]^{-1}.\label{teeresult}
\end{equation}

\subsection{Nonideal interfaces}
\label{tnonideal}

For nonideal interfaces we have to consider both the transmission amplitude $t_{ee}$ from electron to electron and the transmission amplitude $t_{he}$ from electron to hole. It is convenient to define the $2\times 2$ transmission matrix
\begin{equation}
t=\begin{pmatrix}
t_{ee}&t_{eh}\\
t_{he}&t_{hh}
\end{pmatrix},\label{tdef}
\end{equation}
which contains also the transmission amplitudes $t_{eh}$ and $t_{hh}$ from hole to electron and from hole to hole. This matrix $t$ is a $2\times 2$ subblock of the $4\times 4$ unitary scattering matrix $S$, which we derive in Appendix\ \ref{fullS}.

To obtain $t$ from the $4\times 4$ transfer matrix ${\cal M}$, we make a change of basis from the basis $(\Phi_{e,A},\Phi_{h,A},\Phi_{e,C},\Phi_{h,C})$ used in Eq.\ \eqref{Mfull} to a basis $(\Phi_{e,A},\Phi_{h,C},\Phi_{e,C},\Phi_{h,A})$ in which the upper two components are right-movers and the lower two components are left-movers. The change of basis is carried out by the unitary matrix
\begin{equation}
W=\begin{pmatrix}
1&0&0&0\\
0&0&0&1\\
0&0&1&0\\
0&1&0&0
\end{pmatrix}.\label{Wdef}
\end{equation}
We can then follow the same reasoning as in the previous subsection, to conclude that $t$ is determined by the $2\times 2$ upper-left block $X_{11}$ of $W{\cal M}W^{\dagger}$,
\begin{equation}
X_{11}^{\dagger}t=1,\label{tXrelation}
\end{equation}
cf.\ Eq.\ \eqref{tMrelation}.

Substitution of ${\cal M}$ from Eq.\ \eqref{Mfull} gives, after some algebra,
\begin{widetext}
\begin{align}
t^{\dagger}={}&\frac{(\Gamma_1\Gamma_2)^{1/2}}{Z}\begin{pmatrix}
(\MM_{C})_{22}+e^{i\phi(L)}(\MM_A)_{22}\sqrt{1 - \Gamma_1}\sqrt{1-\Gamma_2}\quad&-(\MM_A)_{12}\sqrt{1-\Gamma_1}-e^{-i\phi(L)}(\MM_{C})_{12}\sqrt{1-\Gamma_2}\\
-(\MM_{C})_{21}\sqrt{1-\Gamma_1}-e^{i\phi(L)}(\MM_A)_{21}\sqrt{1-\Gamma_2}\quad&(\MM_A)_{11}+e^{-i\phi(L)}(\MM_{C})_{11}\sqrt{1-\Gamma_1}\sqrt{1-\Gamma_2}
\end{pmatrix},\label{tresult}\\
Z={}&\sqrt{1-\Gamma_1}\sqrt{1-\Gamma_2}\left(e^{-i\phi(L)}{\rm Det}\,\MM_{C}+e^{i\phi(L)}{\rm Det}\,\MM_A\right)+(\MM_A)_{11}(\MM_{C})_{22}\nonumber\\
&+(\MM_A)_{22}(\MM_{C})_{11}(1-\Gamma_1)(1-\Gamma_2)-(\MM_A)_{12}(\MM_{C})_{21}(1-\Gamma_1)-(\MM_A)_{21}(\MM_{C})_{12}(1-\Gamma_2).\label{Zdef}
\end{align}
\end{widetext}

\section{Electrical current}
\label{electric}

\subsection{Ideal interfaces}
\label{Gideal}

Turning now to observable quantities, we will work in the linear response regime $V\rightarrow 0$, when the transmission amplitudes may be evaluated at the Fermi level ($\varepsilon=0$). 

The current $I_{2}^{A}$ (per layer) transmitted into metal contact ${\rm N}_{2}$ through nodal point $A$ is obtained by integrating the transmission probability $|t_{ee}|^{2}$ over $q_A$,
\begin{equation}
I_{2}^{A}=G_{0}V\frac{W}{2\pi}\int dq_A\,|t_{ee}|^{2}.\label{I2A1}
\end{equation}
(The conductance quantum $G_{0}=2e^{2}/h$ includes a twofold spin degeneracy.) The integrand decays exponentially for $|q_A|\gg v_{\alpha}^{2}/v_{F}v_{\Delta}L\simeq (\xi_0/L)k_{F}$. For $L\gg\xi_0$ the effective integration range is much smaller than $k_{F}$ and may be extended to $\pm\infty$. Substituting Eq.\ \eqref{teeresult} (for $\varepsilon=0$) we arrive at
\begin{equation}
I_{2}^{A}=G_{0}V\frac{W}{L}\frac{v_{\alpha}^{2}}{\pi v_{F}v_{\Delta}}.
\label{I2A2}
\end{equation}
As expected, the conductance of a single nodal point has the same form as that of a single valley in a graphene strip, with $L$ 
replaced by the effective propagation length $L_\alpha$ of Eq.~\eqref{Lalpha_def}.

The current $I_{2}^{B}$ transmitted through nodal point $B$ is given by the same formula with 
$v_\alpha$ replaced by $v_{\alpha-\pi/2}$.
Because of the identity
\begin{equation}
v_{\alpha}^{2}+v_{\alpha-\pi/2}^{2}=v_{F}^{2}+v_{\Delta}^{2},\label{videntity}
\end{equation}
the total current $I_{2}=I_{2}^{A}+I_{2}^{B}$ becomes \emph{independent} of $\alpha$.
The conductivity $\sigma_{\rm ideal}=(I_{2}/V)(L/W)$ per layer for the case of ideal NS interfaces is then equal to
\begin{equation}
\sigma_{\rm ideal}=G_{0}\frac{v_{F}^{2}+v_{\Delta}^{2}}{\pi v_{F}v_{\Delta}}.
\label{sigmaideal}
\end{equation}
As discussed in Sec.\ \ref{bulk}, Eq.\ \eqref{sigmaideal} differs [by a factor $1+(v_{\Delta}/v_{F})^{2}$] from the bulk electrical conductivity of Refs.\ \cite{Lee93,Dur00}.

\subsection{Nonideal interfaces}
\label{Gmisaligned}

For nonideal NS interfaces, tunnel barriers couple the nodal points, and the calculation of the current $I_2$ becomes more involved. In this Section we treat the generic case of misaligned nodal points. The case of (perfectly) aligned nodal points is considered 
in Appendix\ \ref{Galigned}.

We first calculate the current through nodal point $A$. 
As discussed in Sect.~\ref{nonideal}, we can substitute $\MM_C$ with $\MM_{\rm Andreev}$ of Eq.\ \eqref{MClimit}, and   
using Eq.\ \eqref{tresult} we obtain the transmission matrix (at $\varepsilon=0$)
\begin{align}
t^{\dagger}_A={}&\frac{\sqrt{\Gamma_1\Gamma_2}}{e^{i\phi_{\alpha}(q_A)}Z_A}\nonumber\\
&\times\begin{pmatrix}
1& - ie^{-i\phi(L)}\sqrt{1-\Gamma_2}\\
i\sqrt{1-\Gamma_1}&~e^{-i\phi(L)}\sqrt{1-\Gamma_1}\sqrt{1-\Gamma_2}
\end{pmatrix},\label{tresultA}
\end{align}
where the denominator $Z_A$ has the form
\begin{align}
Z_A&=\sqrt{\Gamma_1(2-\Gamma_1)\Gamma_2(2-\Gamma_2)}\cosh [L_\alpha (q_A - q_{\rm peak})].\label{ZdefA2}
\end{align}
Here $L_\alpha$ is the effective propagation length \eqref{Lalpha_def}, while $q_{\rm peak}$ is the transverse wave 
number defined by
\begin{align}
q_{\rm peak} = \frac{1}{2L_\alpha} \ln \left[ \frac{\Gamma_1}{2-\Gamma_1} \, \frac{\Gamma_2}{2-\Gamma_2} \right].
\end{align}

Both $t_{ee}$ and $t_{he}$ are peaked at $q_A=q_{\rm peak}$. 
This peak momentum lies at the nodal point ($q_{\rm peak}=0$) only for ideal interfaces. In the presence of tunnel barriers the sign of $q_{\rm peak}$ is such that the order parameter has opposite sign at the two intersections of the 
line $q_A=q_{\rm peak}$ with the Fermi surface.

Integration over $q_A$ of electron current minus hole current gives the net (electrical) current,
\begin{align}
I_{2}^{A}&=G_{0}V\frac{W}{2\pi}\int_{-\infty}^{\infty} dq_A\,\bigl[|(t_A)_{ee}|^{2}-|(t_A)_{he}|^{2}\bigr]\nonumber\\
&=G_{0}V\frac{W}{L}\frac{v_{\alpha}^{2}}{\pi v_{F}v_{\Delta}}\frac{1}{2-\Gamma_1} \frac{\Gamma_2}{2-\Gamma_2}.\label{I2Anonideal}
\end{align}

Similarly, for the current through nodal point $C$ 
we take the limit $q_A\rightarrow \infty$ of Eq.\ \eqref{tresult} and obtain the transmission matrix
 \begin{align}
t^{\dagger}_{C}={}&\frac{\sqrt{\Gamma_1\Gamma_2}}{e^{i\phi_{\alpha}(q_{C})}Z_{C}}\nonumber\\
&\times\begin{pmatrix}
e^{i\phi(L)}\sqrt{1-\Gamma_1}\sqrt{1-\Gamma_2}& i\sqrt{1-\Gamma_1}\\
- ie^{i\phi(L)}\sqrt{1-\Gamma_2}&1
\end{pmatrix},\label{tresultC}\\
Z_{C}=&\sqrt{\Gamma_1(2-\Gamma_1) \Gamma_2 (2-\Gamma_2)} \cosh [L_\alpha (q_A+q_{\rm peak})],
\label{ZdefC}
\end{align}
and then the current
\begin{align}
I_{2}^{C}
&=-G_{0}V\frac{W}{L}\frac{v_{\alpha}^{2}}{\pi v_{F}v_{\Delta}} \frac{1-\Gamma_1}{2-\Gamma_1} \frac{\Gamma_2}{2-\Gamma_2}.\label{I2Cnonideal}
\end{align}
Note the minus sign in the formula for $I_2^{C}$. The current has opposite sign to that at nodal point $A$, since here holes rather than electrons tunnel across the system to contact 2. 

The total current (per layer) through nodal points $A$ and $C$ becomes
\begin{equation}
I_{2}^{A}+I_{2}^{C}=G_{0}V\frac{W}{L}\frac{v_{\alpha}^{2}}{\pi v_{F}v_{\Delta}} \frac{\Gamma_1}{2-\Gamma_1} \frac{\Gamma_2}{2-\Gamma_2}.\label{I2ACnonideal}
\end{equation}
Comparison with Eq.~\eqref{I2A2} reveals that each tunnel barrier changes the sum of the current transmitted through a nodal point and the one opposite to it in momentum space, its time-reversed partner, by a factor of $\Gamma/(2-\Gamma)$.    
As in the case of ideal NS contacts, the pair of nodal points $B$ and $D$ contribute a same amount, but with $v_{\alpha}$ replaced by $v_{\pi/2-\alpha}$. The $\alpha$-dependence again drops out of the total current 
$I_{2}=I_{2}^{A}+ I_2^B + I_2^C + I_2^D$. For the conductivity $\sigma=(I_{2}/V)(L/W)$ per layer we finally obtain
\begin{equation}
\sigma=\sigma_{\rm ideal} \frac{\Gamma_1}{2-\Gamma_1} \frac{\Gamma_2}{2-\Gamma_2}.
\label{sigmanonideal}
\end{equation}

\section{Thermal current}
\label{thermal}

The conductivity \eqref{sigmanonideal} vanishes in the weak tunneling limit $\Gamma_1,\Gamma_2\rightarrow 0$, because the electron and hole contributions to the electrical current $I_{2}$ then become equal but of {\em opposite sign}. Electrons and holes contribute with the {\em same sign\/} to the thermal current,
\begin{equation}
I_{\rm thermal}=L_{0}G_{0}T\delta T\frac{W}{2\pi}\int_{-\infty}^{\infty} dq\,\bigl[|t_{ee}|^{2}+|t_{he}|^{2}\bigr],\label{Ithermal}
\end{equation}
with $L_{0}=\pi^{2}k_{B}^{2}/3e^{2}$ the Lorenz number. The thermal current flows from contact ${\rm N}_{1}$ at temperature $T+\delta T$ into contact ${\rm N}_{2}$ at temperature $T$. (Eq.\ \eqref{Ithermal} requires $\delta T\ll T$ and $T$ sufficiently small that the transmission amplitudes may be evaluated at the Fermi energy $\varepsilon=0$.)

We consider the (generic) case of misaligned nodes. Substitution of the expressions for $t$ from Sec.\ \ref{Gmisaligned}, and summing over the pair of nodal points $A$ and $C$, we find that 
\begin{equation} 
I_{\rm thermal}^{A} +I_{\rm thermal}^{C} =L_{0}G_{0}T\delta T\frac{W}{L}\frac{v_\alpha^2}{\pi v_F v_\Delta}.
\label{Ithermalresult}
\end{equation} 
Quite surprisingly, this turns out to be {\em independent of the tunnel probabilities\/} $\Gamma_1$ and $\Gamma_2$.
The total thermal current (per layer) also includes contributions from the nodal points $B$ and $D$, and is -- just as the electrical conductivity -- independent of the angle $\alpha$: 
\begin{equation}
I_{\rm thermal} = L_{0}G_{0}T\delta T\frac{W}{L}\frac{v_F^2 + v_\Delta^2}{\pi v_F v_\Delta} .
\label{Ithermalresult}
\end{equation}
 As discussed in Sec.\ \ref{bulk}, the thermal conductivity $\kappa=(I_{\rm thermal}/\delta T)(L/W)$ extracted from Eq.\ \eqref{Ithermalresult}  coincides with the bulk thermal conductivity of Ref.\ \cite{Dur00}.

\section{Shot noise}
\label{shotnoise}

The zero-frequency noise power of time dependent electrical current fluctuations $\delta I_{2}(t)$ measured in contact number 2,
\begin{equation}
P_{22}=\int_{-\infty}^{\infty}dt\,\overline{\delta I_{2}(0)\delta I_{2}(t)},\label{P2def}
\end{equation}
is given in terms of the transmission matrix elements by the general expression \cite{Ana96}
\begin{align}
P_{22}={}&G_{0}eV\frac{W}{2\pi}\int dq\,\bigl[|t_{ee}|^{2}(1-|t_{ee}|^{2})\nonumber\\
&+|t_{he}|^{2}(1-|t_{he}|^{2})+2|t_{he}|^{2}|t_{ee}|^{2}\bigr].\label{P22}
\end{align}
As with the conductance, we work in the linear response regime, so the transmission matrix is to be evaluated at $\varepsilon=0$.

We restrict ourselves to the case of misaligned nodes and substitute the expressions for $t$ from Sec.\ \ref{Gmisaligned}. The integral over $q$ contains four separate contributions, from $q$ near nodes $A$, $B$, $C$, and $D$. The total result (per layer) is
\begin{align}
P_{22}={}&G_{0}eV\frac{W}{L}\frac{v_{F}^{2}+v_{\Delta}^{2}}{\pi v_{F}v_{\Delta}}\nonumber\\
&\times\frac{12(2-\Gamma_1)^{2}(1-\Gamma_2)+8(1-\Gamma_1)\Gamma_2^{2}+\Gamma_1^{2}\Gamma_2^{2}}{3(2-\Gamma_1)^{2}(2-\Gamma_2)^{2}}.\label{P22result}
\end{align}
The Fano Factor $F=P_{22}/eI_{2}$ is given by
\begin{equation}
F=\frac{12(2-\Gamma_1)^{2}(1-\Gamma_2)+8(1-\Gamma_1)\Gamma_2^{2}+\Gamma_1^{2}\Gamma_2^{2}}{3\Gamma_1\Gamma_2(2-\Gamma_1)(2-\Gamma_2)}.\label{Fano}
\end{equation}

In the ideal limit $\Gamma_1,\Gamma_2\rightarrow 1$ we find a Fano factor $F=1/3$, three times smaller than the value $F=1$ associated with a Poisson process. As discussed in the context of graphene \cite{Two06,DiC08,Dan08}, this is the same one-third reduction as in a diffusive metallic conductor and is a hallmark of pseudodiffusive transmission.

In the weak tunneling limit $\Gamma_1,\Gamma_2\rightarrow 0$ the noise power remains finite,
\begin{equation}
\lim_{\Gamma_1,\Gamma_2\rightarrow 0}P_{22}=G_{0}eV\frac{W}{L}\frac{v_{F}^{2}+v_{\Delta}^{2}}{\pi v_{F}v_{\Delta}},\label{P22weak}
\end{equation}
while the electrical current vanishes, $I_{2}\propto\Gamma_1\Gamma_2\rightarrow 0$. The electrical current fluctuations therefore become large relative to the time-averaged current in the presence of tunnel barriers. This is discussed in the context of resonant tunneling through midgap states in Sec.\ \ref{midgap}.

\section{Discussion}
\label{discuss}

\subsection{Comparison with bulk electrical and thermal conductivities}
\label{bulk}

The electrical current $I_{2}$ and thermal current $I_{\rm thermal}$ that we have calculated describe transmission of electrons and holes over a finite length $L$ of a clean \textit{d}-wave superconductor. Earlier work \cite{Lee93,Dur00} calculated the electrical and thermal conductivities $\sigma_{0}$ and $\kappa_{0}$ of a disordered infinite system. These are in principle different systems, but we can still compare them by formally converting the currents through the finite system into bulk conductivities by means of $\sigma_{0}\equiv (I_{2}/V)(L/W)$ and $\kappa_{0}\equiv (I_{\rm thermal}/\delta T)(L/W)$. 

The thermal conductivity obtained in this way from the finite-system thermal current \eqref{Ithermalresult},
\begin{equation}
\kappa_{0}=L_0 G_{0}T\frac{v_{F}^{2}+v_{\Delta}^{2}}{\pi v_{F}v_{\Delta}},\label{sigmathermal}
\end{equation}
is the same as the bulk thermal conductivity of Durst and Lee \cite{Dur00}. The results for the electrical conductivity differ, however. The bulk result \cite{Lee93,Dur00}
\begin{equation}
\sigma_{0}=G_{0}\frac{v_{F}}{\pi v_{\Delta}}\label{sigmaelectric}
\end{equation}
differs from the finite-system result \eqref{sigmaideal} --- even if we assume ideal NS interfaces. The difference between the factors $v_{F}/v_{\Delta}$ in Eq.\ \eqref{sigmaelectric} and $(v_{F}^{2}+v_{\Delta}^{2})/v_{F}v_{\Delta}$ in Eq.\ \eqref{sigmaideal} is small in practice (because $v_{F}\gg v_{\Delta}$), but the difference does illustrate that these are different systems.

\subsection{Interpretation in terms of resonant tunneling through midgap states}
\label{midgap}

We have found that tunnel barriers at the NS interfaces reduce the transmitted electrical current, but not the thermal current nor the electrical noise. This result has a natural interpretation in terms of the midgap states at the NS interfaces. Midgap states are zero-energy edge states of the \textit{d}-wave superconductor, which exist at momentum $q$ along the edge if the order parameter has opposite sign at the two intersections of the line of constant $q$ with the Fermi surface \cite{Hu94,Kas00}. The midgap states at the two NS interfaces have a small overlap, and therefore acquire a nonzero energy $\pm E_{\rm edge}$ (tunnel splitting). Moreover, the coupling to the metal electrodes at $s=0,L$ introduces partial widths $\delta E_{0}$, $\delta E_{L}$ of the midgap states (tunnel broadening). 

Tunneling through a pair of midgap states was studied in Ref.\ \cite{Nil08}, in the context of Majorana bound states (which are a special type of nondegenerate midgap states). We can compare the transmission probabilities resulting from that work,
\begin{equation}
|t_{ee}|^{2}=|t_{eh}|^{2}=|t_{he}|^{2}=|t_{hh}|^{2}=\frac{E_{\rm edge}^{2}\delta E_{0}\delta E_{F}}{(E_{\rm edge}^{2}+\delta E_{0}\delta E_{L})^{2}},\label{tmajorana}
\end{equation}
with the results from Sec.\ \ref{Gmisaligned} in the tunneling limit $\Gamma_1,\Gamma_2\ll 1$,
\begin{align}
|t_{ee}|^{2} &=|t_{eh}|^{2}=|t_{he}|^{2}=|t_{hh}|^{2}=\frac{1}{4\cosh^{2}[L_\alpha(q-q_{\rm peak})]}.
\label{tAtC}
\end{align}
We have defined 
\begin{align}
e^{L_\alpha q_{\rm peak}} &=\tfrac{1}{2}\sqrt{\Gamma_1\Gamma_2}.\label{xi0}
\end{align}
This comparison leads to the identification
\begin{equation}
\frac{E_{\rm edge}}{\sqrt{\delta E_{0}\delta E_{L}}}=\frac{2e^{L_\alpha q}}{\sqrt{\Gamma_1\Gamma_2}}.\label{identify}
\end{equation}

Resonant tunneling, with all transmission probabilities equal to $1/4$, occurs when $q=q_{\rm peak}$, hence when $E_{\rm edge}=\sqrt{\delta E_{0}\delta E_{L}}$ (tunnel splitting of the midgap states equal to tunnel broadening). Because transmission from electron to electron and from electron to hole happens with the same probability (to leading order in $\Gamma_1,\Gamma_2$), the transmitted electrical current vanishes in the limit of small $\Gamma$. The thermal current $I_{\rm thermal}$ and electrical noise $P_{22}$ remain finite, because $|t_{ee}|^{2}$ and $|t_{he}|^{2}$ contribute with the same sign to these quantities.

This interpretation explains the finite small-$\Gamma$ limit for $P_{22}$ and $I_{\rm thermal}$, but it does not explain why the thermal current \eqref{Ithermalresult} turns out to be completely independent on the values of $\Gamma_1$ and $\Gamma_2$. That remains a surprising result of our calculation, for which we have no qualitative explanation.

\subsection{Outlook}
\label{outlook}

We have shown how ballistic transport through a clean \textit{d}-wave superconductor (such as single-crystal ${\rm YBa}_{2}{\rm Cu}_{3}{\rm O}_{7-\epsilon}$) has features in common with graphene \cite{Bee08}: a pseudodiffusive $1/L$ scaling of the electrical current transmitted over a distance $L$, and a $1/3$ suppresssion of the electrical shot noise with respect to the Poisson value of uncorrelated current pulses. These effects have been observed in graphene \cite{Mia07,DiC08,Dan08} and it would be interesting to search for them in the high-$T_{c}$ cuprates. The $1/L$ scaling should persist, with a modified slope, in the presence of tunnel barriers at the NS interfaces, and in the case of the thermal current we find that even the slope is independent of the tunnel barrier height.

There are more areas of correspondence between massless Dirac fermions in \textit{d}-wave superconductors and in graphene, in addition to the pseudodiffusive transport studied in this work. We mention two such effects, as directions for future research.
\begin{itemize}
\item
In graphene an electrostatic potential can displace the Fermi level away from the Dirac point of vanishing density of states. In the \textit{d}-wave superconductor the supercurrent velocity $\bm{v}_{s}$ enters into the Dirac equation \eqref{Dirac2} as a scalar term $\propto\sigma_{0}$ \cite{Fra00}, and therefore has the same effect of displacing the Dirac point relative to the Fermi level. There is one curious difference with respect to graphene: the \textit{d}-wave superconductor has two pairs of valleys and the Dirac point can be displaced independently  in each pair (relative to the same Fermi level). With reference to Fig.\ \ref{fig_Brillouin}, the component of $\bm{v}_{s}$ in the $x$-direction acts on valleys at the nodal points $A$ and $C$, while the component in the $y$-direction acts on those at $B$ and $D$.
\item
While the role of an electrostatic potential in graphene is played by the supercurrent, an electric field in the \textit{d}-wave superconductor plays the role of a magnetic field in graphene. If a sufficiently strong electric field could be induced in a thin-film cuprate superconductor, it might be possible to see effects analogous to the effects of Landau level quantization in graphene \cite{Cas09}.
\end{itemize}

\acknowledgments

This research was supported by the Dutch Science Foundation NWO/FOM and by an ERC Advanced Investigator Grant.

\appendix

\section{NS interface matrix}
\label{MNS_derivation}

In Sec.~\ref{tmideal} we derived the most general form of the transfer matrix of an NS interface, consistent with the requirement of particle flux conservation. The result in Eq.\ \eqref{MNSdef} has three undetermined parameters $\beta_x$, $\beta_y$, and $\beta_z$. Here we calculate the interface matrix by solving the Dirac equation in the interface layer and determine these unknown parameters. The interface layer is the region where the order parameter increases from $0$ to its bulk value, over a healing length $l_0$ 
(which is typically of the same order of magnitude as the coherence length $\xi_0$). 

As discussed in Ref.~\cite{Sim97}, in order to preserve Hermiticity, the Dirac equation \eqref{Dirac2}
needs to be supplemented by terms containing the spatial derivatives of $v_{\Delta}$:
\begin{equation}
-i\left[v_F \sigma_z \partial_x + v_\Delta \sigma_x \partial_y + (\partial_yv_\Delta)\sigma_x/2 \right] \Psi 
= \varepsilon \Psi.
\label{Dirac4}
\end{equation}
We assumed that the phase of $\Delta$ is constant, and set it to 0 (without loss of generality), thus $v_\Delta$ is real throughout, 
and is only a function of the distance $s$ from the NS interface.  
An eigenstate $\Psi(s)$ of momentum $q$ parallel to the NS interface satisfies
\begin{equation}
\left[ -i J \partial_s + J_1 q -i v_\Delta'  \sin(\alpha) \sigma_x /2 \right] \Psi(s) = \varepsilon \Psi(s),
\end{equation}
with the derivative of $v_\Delta$ denoted by the shorthand $v_\Delta' \equiv \partial_s v_\Delta$.
Accordingly, the matrix $\aaa$ in Eq.~\eqref{diff_A} becomes $s$-dependent and gets a new term: 
\begin{align} 
\aaa(s) &= 
\aaa_0(s) \nonumber \\
&- \frac{v_\Delta'(s) \sin \alpha }{2 v_\alpha(s)^2} \left( \sigma_y v_F \cos\alpha -i v_\Delta(s) \sin \alpha \right).
\end{align} 

Since $ql_{0}v_{F}v_{\Delta}/v_{\alpha}^{2}\simeq q/k_{F}\ll 1$ (in the relevant range of $q$'s near the nodal point), the integral of ${\cal A}_{0}$ over the interface layer is $\ll 1$ and may be neglected.
Then $\aaa(s_1)$ commutes with $\aaa(s_2)$ for $0<s_1,s_2<l_0$, and therefore 
we can simply integrate Eq.~\eqref{diff_A} over the interface layer:
\begin{align}
\MM_{NS}& = \exp\left[i\int_0^{l_0} \aaa(s) ds\right] \nonumber\\
\quad &= \exp\left[\frac{-i}{2}\int_0^{\tan \theta} \frac{1}{1+u^2} \left( \sigma_y - i u \right) du\right] \nonumber\\
\quad &= \sqrt{\frac{v_F \cos \alpha }{ v_\alpha}} \exp[-i\theta \sigma_y /2],
\end{align}
with 
$\theta = \arctan \left[ (v_\Delta/v_F) \tan \alpha  \right]$ as defined in Eq.~\eqref{theta_def}. 
The result agrees with Eq.~\eqref{MNSdef} with $\beta_x=\beta_y=\beta_z=0$, and $v_N=v_F \cos \alpha$.
The Fermi velocity mismatch contributes an additional factor $\sqrt{v_{N}/v_{F}}$ to the interface matrix, and in addition may cause internode scattering (as detailed in Sec.\ \ref{nonideal}).

\section{Full scattering matrix}
\label{fullS}

In Secs.\ \ref{transmission} and \ref{electric} we have calculated the $2\times 2$ transmission matrix $t$, which is the quantity we need for the transport properties considered. For reference, we give here the full $4\times 4$ scattering matrix,
\begin{equation}
S=\begin{pmatrix}
r&t'\\
t&r'
\end{pmatrix},\label{Sdef}
\end{equation}
containing the $2\times 2$ transmission matrices $t$ (from left to right) and $t'$ (from right to left), as well as the reflection matrices $r$ (from left to left) and $r'$ (from right to right). These matrices can be obtained from transfer matrix ${\cal M}$ by constructing the four $2\times 2$ sub-blocks $X_{ij}$,
\begin{equation}
W{\cal M}W^{\dagger}=\begin{pmatrix}
X_{11}&X_{12}\\
X_{21}&X_{22}
\end{pmatrix},\label{WMWXrelation}
\end{equation}
and then evaluating
\begin{align}
&r=-X_{22}^{-1}X_{21},\;\;
r'=X_{12}X_{22}^{-1},
\nonumber\\
&t^{\dagger}=X_{11}^{-1},\;\;t'=X_{22}^{-1},\label{SMXrelation}
\end{align}
cf.\ Eqs.\ \eqref{Wdef} and \eqref{tXrelation}.

We restrict ourselves to $\varepsilon=0$ and misaligned nodes. Near node $A$ we find the reflection matrices
\begin{widetext}
\begin{align}
r_A&=\frac{1}{Z_A}\begin{pmatrix}
-e^{ L_\alpha q_A}\sqrt{1-\Gamma_1}(2-\Gamma_2)&
~ - i\Gamma_1(e^{ L_\alpha q_A} - \Gamma_2\sinh L_\alpha q_A)\\
- i\Gamma_1(e^{ L_\alpha q_A}-\Gamma_2\cosh L_\alpha q_A)&
~-e^{ L_\alpha q_A}\sqrt{1-\Gamma_1}(2-\Gamma_2)
\end{pmatrix},\label{rresultA}\\
r'_A&=\frac{1}{Z_A}\begin{pmatrix}
e^{-i\phi(L)}e^{ L_\alpha q_A}\sqrt{1-\Gamma_2}(2-\Gamma_1)&
~- i\Gamma_2(e^{ L_\alpha q_A} - \Gamma_1\cosh L_\alpha q_A)\\
- i\Gamma_2(e^{ L_\alpha q_A} - \Gamma_1\sinh L_\alpha q_A)&
~e^{i\phi(L)}e^{ L_\alpha q_A}\sqrt{1-\Gamma_2}(2-\Gamma_1)
\end{pmatrix}.\label{rprimeresultA}
\end{align}
The transmission matrix $t_A$ is given by Eq.\ \eqref{tresultA} and $t'_A=\sigma_{y}t_A^{\dagger}\sigma_{y}$. The resulting scattering matrix \eqref{Sdef} is unitary, $SS^{\dagger}=1$, as it should be.

Similarly, near node $C$ we find $t_{C}$ given by Eq.\ \eqref{tresultC}, $t'_{C}=\sigma_{y}t_{C}^{\dagger}\sigma_{y}$, and the reflection matrices
\begin{align}
r_{C}&=\frac{1}{Z_{C}}\begin{pmatrix}
-e^{- L_\alpha q_C}\sqrt{1-\Gamma_1}(2-\Gamma_2)&
~- i\Gamma_1(e^{- L_\alpha q_C}-\Gamma_2\cosh L_\alpha q_C)\\
- i\Gamma_1(e^{- L_\alpha q_C} + \Gamma_2\sinh L_\alpha q_C)&
~-e^{- L_\alpha q_C}\sqrt{1-\Gamma_1}(2-\Gamma_2)
\end{pmatrix},\label{rresultC}\\
r'_C&=\frac{1}{Z_{C}}\begin{pmatrix}
e^{-i\phi(L)}e^{- L_\alpha q_C}\sqrt{1-\Gamma_2}(2-\Gamma_1)&~
- i\Gamma_2(e^{- L_\alpha q_C} + \Gamma_1\sinh L_\alpha q_C)\\
- i\Gamma_2(e^{- L_\alpha q_C} - \Gamma_1\cosh L_\alpha q_C)&~
e^{i\phi(L)}e^{- L_\alpha q_C}\sqrt{1-\Gamma_2}(2-\Gamma_1)
\end{pmatrix}.\label{rprimeresultC}
\end{align}
\end{widetext}

\section{Conductance for aligned nodal points}
\label{Galigned}

\subsection{Alignment of nodes $\bm{A-C}$.}
\label{ACalignment}

For $|\alpha|\ll \xi_0/L $ the two nodal points $A$ and $C$ line up with the normal to the NS interface, while nodes $B$ and $D$ remain misaligned. Restricting ourselves again to $\varepsilon=0$, we may put $q_A=q_{C}=q$, $\phi_{\alpha}=0$, $\MM_A=\MM_{C}$, and $L_\alpha = L v_\Delta/v_F \equiv L_0$ in Eq.\ \eqref{tresult}. The result is
\begin{widetext}
\begin{align}
t_{AC}^{\dagger}={}& \frac{(\Gamma_1\Gamma_2)^{1/2}}{Z_{AC}}
\begin{pmatrix}
[1+e^{2ik_{F}L}\sqrt{1 - \Gamma_1}\sqrt{1-\Gamma_2}]\cosh L_0 q\quad 
&i[\sqrt{1-\Gamma_1}+e^{-2ik_{F}L}\sqrt{1-\Gamma_2}]\sinh L_0 q\\
-i[\sqrt{1-\Gamma_1}+e^{2ik_{F}L}\sqrt{1-\Gamma_2}]\sinh L_0 q\quad 
&[1+e^{-2ik_{F}L}\sqrt{1-\Gamma_1}\sqrt{1-\Gamma_2}]\cosh L_0 q
\end{pmatrix},\label{talignedresult}\\
Z_{AC}={}&2\sqrt{1-\Gamma_1}\sqrt{1-\Gamma_2}\cos(2k_{F}L)+2-\Gamma_1-\Gamma_2+\Gamma_1\Gamma_2\cosh^{2} L_0 q.\label{Zaligneddef}
\end{align}

The current $I_{2}^{AC}$ through the aligned nodes $A$ and $C$ follows from
\begin{align}
&I_{2}^{AC}=G_{0}V \frac{W}{2\pi }\int_{-\infty}^{\infty}dq \,(|t_{ee}|^{2}-|t_{he}^{2}|),\label{I2ACaligned}\\
&|t_{ee}|^{2}-|t_{he}^{2}|=\frac{2\Gamma_1\Gamma_2}{(2-\Gamma_1)(2-\Gamma_2)+4\sqrt{1-\Gamma_1}\sqrt{1-\Gamma_2}\cos(2k_{F}L)+\Gamma_1\Gamma_2\cosh (2 L_0 q)}.\label{teetheAC}
\end{align}
\end{widetext}
For the total current $I_{2}$ we add the contribution from the (strongly) misaligned nodes $B$ and $D$,
\begin{equation}
I_{2}=I_{2}^{AC}+G_{0}V\frac{W}{L}\frac{v_{\Delta}}{\pi v_{F}}\frac{\Gamma_1}{2-\Gamma_1} \, \frac{\Gamma_2}{2-\Gamma_2}.\label{I2AC}
\end{equation}

\begin{figure}[tb]
\centerline{\includegraphics[width=0.8\linewidth]{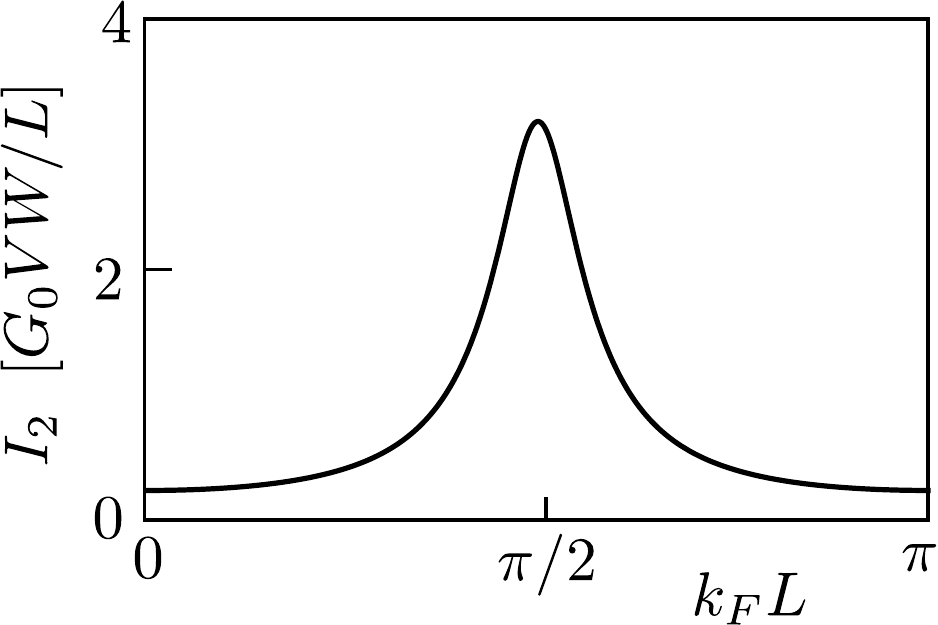}}
\caption{\label{fig_aligned_a}
Dependence on the separation $L$ of the NS interfaces of the current $I_{2}$ into contact ${\rm N}_{2}$, for the interface orientation $\alpha=0$ of aligned nodes $A$ and $C$. Calculated from Eqs.\ \eqref{I2ACaligned}--\eqref{I2AC} for parameters $\Gamma_1=\Gamma_2=0.3$, $v_{F}/v_{\Delta}=10$.
}
\end{figure}

As shown in Fig.\ \ref{fig_aligned_a}, the current $I_{2}$ oscillates as a function of $k_{F}L$, between minima $I_{2}^{\rm min}$ at $k_{F}L=0\,({\rm mod}\,\pi)$ and maxima $I_{2}^{\rm max}$ at $k_{F}L=\pi/2\,({\rm mod}\,\pi)$. (Similar oscillations were found in Ref.\ \cite{Don04}.) Simple expressions for these two values follow for the case $\Gamma_1=\Gamma_2\equiv\Gamma$ of equal tunnel barriers,
\begin{align}
&I_{2}^{\rm min}=G_{0}V\frac{W}{L}\frac{\Gamma^{2}}{\pi(2-\Gamma)^{2}}\left(\frac{v_{F}}{v_{\Delta}}\frac{{\rm artanh}\,\chi}{\chi}+\frac{v_{\Delta}}{v_{F}}\right), \label{I2min}\\
&I_{2}^{\rm max}=G_{0}V\frac{W}{L}\frac{1}{\pi}\left(\frac{v_{F}}{v_{\Delta}}+\frac{v_{\Delta}}{v_{F}}\frac{\Gamma^{2}}{(2-\Gamma)^{2}}\right), \label{I2max}
\end{align}
with abbreviation $\chi=2(2-\Gamma)^{-1}\sqrt{1-\Gamma}$. For $\Gamma=1$ we recover the ideal limit $I_{2}^{\rm min}=I_{2}^{\rm max}=\sigma_{\rm ideal}VW/L$. For $\Gamma\ll 1$ we have instead $I_{2}^{\rm max}=G_{0}V(W/\pi L)(v_{F}/v_{\Delta})$, $I_{2}^{\rm min}=I_{2}^{\rm max}\times\tfrac{1}{4}\Gamma^{2}|\ln \Gamma|$.

\subsection{Alignment of nodes $\bm{A-D}$ and $\bm{B-C}$}
\label{AD_BC_alignment}

\begin{figure}[tb]
\centerline{\includegraphics[width=0.7\linewidth]{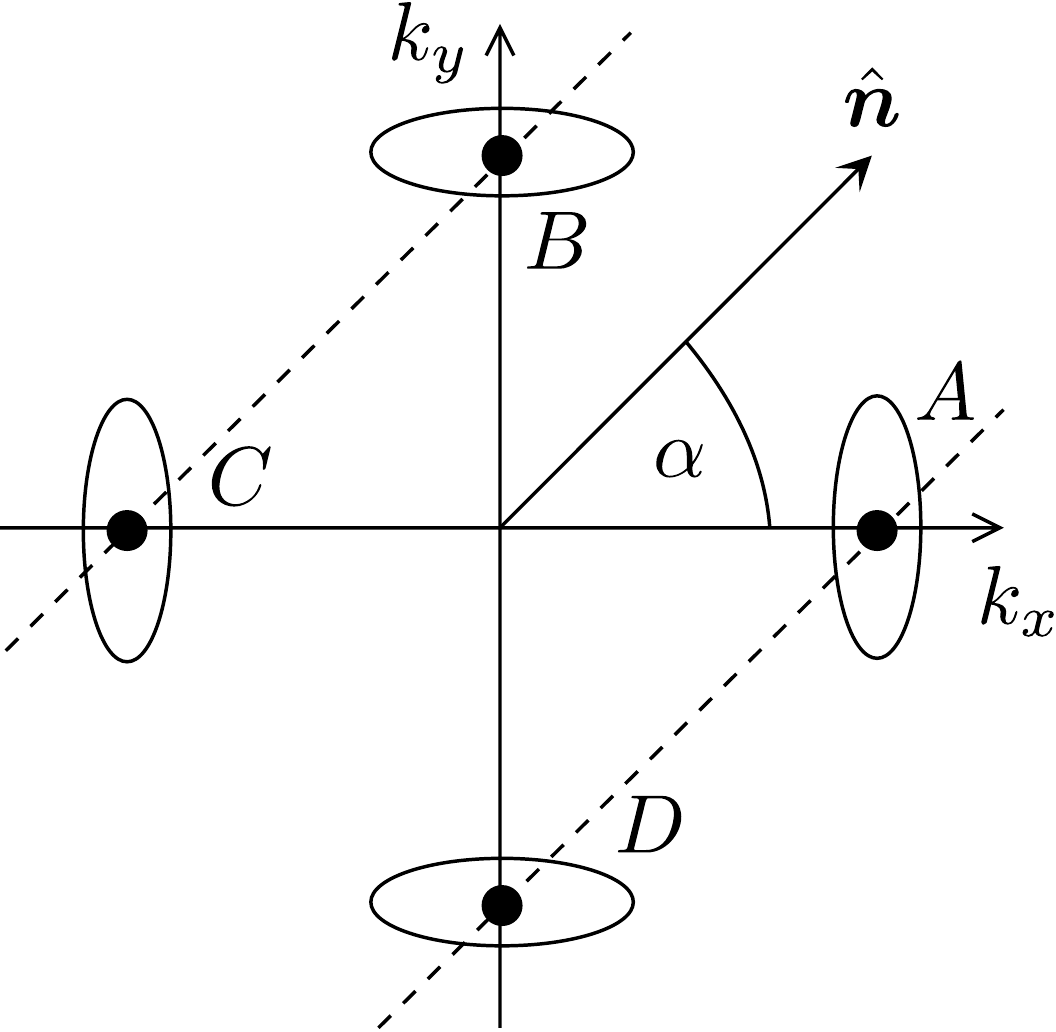}}
\caption{\label{fig_Nodes}
Same as Fig.\ \ref{fig_Brillouin}, but now for an angle $\alpha=\pi/4$ between the normal $\bm{\hat{n}}$ to the NS interface and the lines $x=0$, $y=0$ of vanishing order parameter. For this orientation the nodal points $A-D$ and $B-C$ are pairwise aligned with $\bm{\hat{n}}$ (dashed lines), so that they are pairwise coupled by a tunnel barrier at the interfaces.
}
\end{figure}

For $|\alpha-\pi/4| \ll \xi_0/L$, nodal points $A-D$ and $B-C$ are pairwise aligned with the normal to the NS interface (see Fig.\ \ref{fig_Nodes}). The transmission matrix $t_{AD}$ through nodes $A-D$ is given by Eqs.\ \eqref{tresult} and \eqref{Zdef} with $\MM_{C}$ replaced by $\MM_{D}$. Similarly, for the transmission matrix $t_{BC}$ through nodes $B-C$ we should replace $\MM_A$ by $\MM_{B}$.

Considering first the transmission through nodes $A-D$, we see from Eq.\ \eqref{MABCD} that $\MM_{D}=\MM_A^{-1}$ at $|\alpha|=\pi/4$, $q_A=q_{D}\equiv q$. Restricting ourselves to $\varepsilon=0$, we find
\begin{widetext}
\begin{align}
t_{AD}^{\dagger}={}&\frac{2(\Gamma_1\Gamma_2)^{1/2}}{e^{i\phi_{\pi/4}(q_A)}Z_{AD}}\begin{pmatrix}
\bigl(1+e^{i\psi( q)}\sqrt{1-\Gamma_1}\sqrt{1-\Gamma_2}\bigr)\cosh  L_\alpha q
&-ie^{-ik_{F}L\sqrt{2}}\bigl(\sqrt{1-\Gamma_2}-e^{i\psi(q)}\sqrt{1-\Gamma_1}\bigr)\sinh  L_\alpha q\\
-i\bigl(e^{i\psi(q)}\sqrt{1-\Gamma_2}-\sqrt{1-\Gamma_1}\bigr)\sinh  L_\alpha q
&e^{-ik_{F}L\sqrt{2}}\bigl(e^{i\psi(q)}+\sqrt{1-\Gamma_1}\sqrt{1-\Gamma_2}\bigr)\cosh  L_\alpha q
\end{pmatrix},\label{tADresult}\\
Z_{AD}={}&\Gamma_1\Gamma_2+4\sqrt{1-\Gamma_1}\sqrt{1-\Gamma_2}\cos\psi(q)+(2-\Gamma_1)(2-\Gamma_2)\cosh 2 L_\alpha q,\label{ZADdef}
\end{align}
with $L_\alpha = 2 L v_F v_\Delta / (v_F^2 + v_\Delta^2) $
and the auxiliary function $\psi(q)=2 L q (v_{F}^{2}-v_{\Delta}^{2})/ (v_{F}^2 + v_\Delta^2) + k_{F} L\sqrt{2}$.

The current $I_{2}^{AD}$ through the aligned nodes $A$ and $D$ follows from
\begin{align}
&I_{2}^{AD}=G_{0}V\frac{W}{2\pi}\int_{-\infty}^{\infty}dq\,(|t_{ee}|^{2}-|t_{he}^{2}|),\label{I2ADaligned}\\
&|t_{ee}|^{2}-|t_{he}^{2}|
=2\Gamma_1\Gamma_2
\frac{(2-\Gamma_1)(2-\Gamma_2)
+\bigl[\Gamma_1\Gamma_2+4\sqrt{1-\Gamma_1}\sqrt{1-\Gamma_2}\cos\psi(q)\bigr]
\cosh(2 L_\alpha q)}
{\bigl[\Gamma_1\Gamma_2+4\sqrt{1-\Gamma_1}\sqrt{1-\Gamma_2}\cos\psi(q)+(2-\Gamma_1)(2-\Gamma_2)
\cosh (2 L_\alpha q)\bigr]^{2}}
.\label{teetheAD}
\end{align}
\end{widetext}
The contribution from the aligned nodes $B$ and $C$ is identical, so the total current becomes $I_{2}=2I_{2}^{AD}$. 

\begin{figure}[tb]
\centerline{\includegraphics[width=0.8\linewidth]{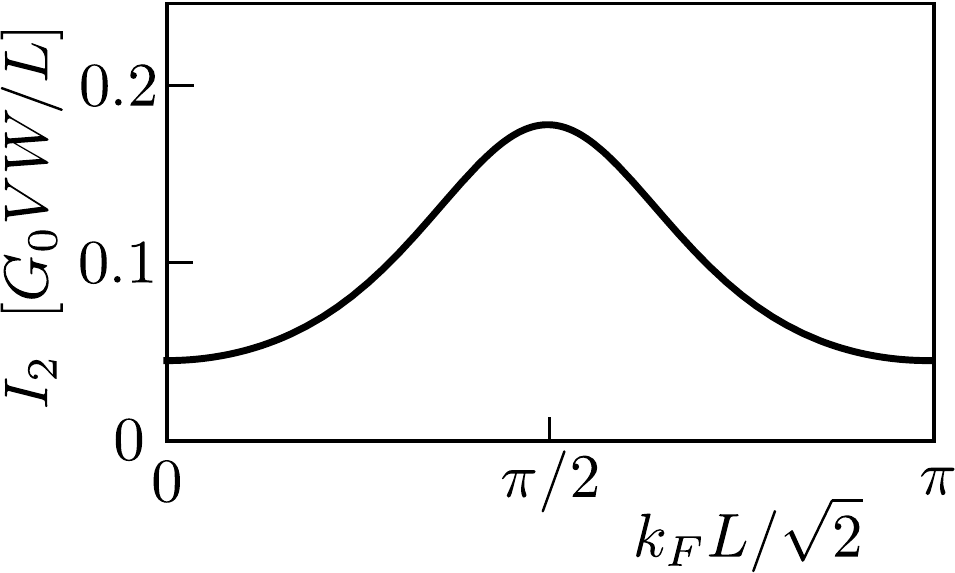}}
\caption{\label{fig_aligned_b}
Same as Fig.\ \ref{fig_aligned_a}, but now for an interface orientation $\alpha=\pi/4$ of pairwise aligned nodes $A-D$ and $B-C$, calculated from Eq.\ \eqref{I2ADaligned} for parameters $\Gamma_1=\Gamma_2=0.3$, $v_{F}/v_{\Delta}=10$.
}
\end{figure}

For ideal interfaces ($\Gamma_1=\Gamma_2=1$), we recover the result $I_{2}=\sigma_{\rm ideal}VW/L$. In the presence of tunnel barriers, $I_{2}$ again oscillates as a function of $L$, see Fig.\ \ref{fig_aligned_b}.


\begin{thebibliography}{99}
\bibitem{Bee08} C. W. J. Beenakker, Rev. Mod. Phys. \textbf{80}, 1337 (2008).
\bibitem{Kat06} M. I. Katsnelson, Eur. Phys. J. B \textbf{51}, 157 (2006).
\bibitem{Two06} J. Tworzyd{\l}o, B. Trauzettel, M. Titov, A. Rycerz, and C. W. J. Beenakker, Phys. Rev. Lett. \textbf{96}, 246802 (2006).
\bibitem{Akh07} A. R. Akhmerov and C. W. J. Beenakker, Phys. Rev. B \textbf{75}, 045426 (2007).
\bibitem{Pra07} E. Prada, P. San-Jose, B. Wunsch, and F. Guinea, Phys. Rev. B \textbf{75}, 113407 (2007).
\bibitem{Sch07} H. Schomerus, Phys. Rev. B \textbf{76}, 045433 (2007).
\bibitem{Bla07} Y. M. Blanter and I. Martin, Phys. Rev. B \textbf{76}, 155433 (2007).
\bibitem{Cse07} J. Cserti, A. Csordas, and G. David, Phys. Rev. Lett. \textbf{99}, 066802 (2007).
\bibitem{Cre07} A. Cresti, G. Grosso, and G. Pastori Parravicini, Phys. Rev. B \textbf{76}, 205433 (2007).
\bibitem{Tit07} M. Titov, EPL \textbf{79}, 17004 (2007).
\bibitem{Mog09} A. G. Moghaddam and M. Zareyan, Phys. Rev. B \textbf{79}, 073401 (2009).
\bibitem{Die09} P. Dietl, G. Metalidis, D. Golubev, P. San-Jose, E. Prada, H. Schomerus, and G. Sch\"{o}n, Phys. Rev. B \textbf{79}, 195413 (2009).
\bibitem{Mia07}F. Miao, S. Wijeratne, Y. Zhang, U. C. Coskun, W. Bao, and C. N. Lau, Science {\bf 317}, 1530 (2007).
\bibitem{DiC08} L. DiCarlo, J. R. Williams, Y. Zhang, D. T. McClure, and C. M. Marcus, Phys. Rev. Lett. \textbf{100}, 156801 (2008).
\bibitem{Dan08} R. Danneau, F. Wu, M. F. Craciun, S. Russo, M. Y. Tomi, J. Salmilehto, A. F. Morpurgo, and P. J. Hakonen, Phys. Rev. Lett. \textbf{100}, 196802 (2008).
\bibitem{Sep07} R. A. Sepkhanov, Ya. B. Bazaliy, and C. W. J. Beenakker, Phys. Rev. A \textbf{75}, 063813 (2007).
\bibitem{Sep08} R. A. Sepkhanov and C. W. J. Beenakker, Opt. Commun. \textbf{281}, 5267 (2008).
\bibitem{Zha08a} X. D. Zhang, Phys. Lett. A \textbf{372}, 3512 (2008).
\bibitem{Zha08b} X. D. Zhang and Z. Y. Liu, Phys. Rev. Lett. \textbf{101}, 264303 (2008).
\bibitem{Har95} D. J. van Harlingen, Rev. Mod. Phys. \textbf{67}, 515 (1995).
\bibitem{Lee93} P. A. Lee, Phys. Rev. Lett. \textbf{71}, 1887 (1993).
\bibitem{Dur00} A. C. Durst and P. A. Lee, Phys. Rev. B \textbf{62}, 1270 (2000).
\bibitem{Alt02} A. Altland, B. D. Simons, and M. R. Zirnbauer, Phys. Rep. \textbf{359}, 283 (2002).
\bibitem{Har06} R. Harris, P. J. Turner, S. Kamal, A. R. Hosseini, P. Dosanjh, G. K. Mullins, J. S. Bobowski, C. P. Bidinosti, D. M. Broun, R. Liang, W. N. Hardy, and D. A. Bonn, Phys. Rev. B \textbf{74}, 104508 (2006).
\bibitem{Don04} Z. C. Dong, Z. M. Zheng, and D. Y. Xing, J. Phys. Cond. Matt. \textbf{16}, 6099 (2004).
\bibitem{Tak06} S. Takahashi, T. Yamashita, and S. Maekawa, J. Phys. Chem. Sol. \textbf{67}, 325 (2006).
\bibitem{Her09} W. J. Herrera, A. Levy Yeyati, and A. Mart\'{i}n-Rodero, Phys. Rev. B \textbf{79}, 014520 (2009).
\bibitem{Hu94} C.-R. Hu, Phys. Rev. Lett. \textbf{72}, 1526 (1994).
\bibitem{Kas00} S. Kashiwaya and Y. Tanaka, Rep. Prog. Phys. \textbf{63}, 1641 (2000).
\bibitem{Ana96} M. P. Anantram and S. Datta, Phys. Rev. B \textbf{53}, 16390 (1996).
\bibitem{Nil08} J. Nilsson, A. R. Akhmerov, and C. W. J. Beenakker, Phys. Rev. Lett. \textbf{101}, 120403 (2008).
\bibitem{Fra00} M. Franz and Z. Te\v{s}anovi\'{c}, Phys. Rev. Lett. \textbf{84}, 554 (2000).
\bibitem{Cas09} A. H. Castro Neto, F. Guinea, N. M. Peres, K. S. Novoselov, and A. K. Geim, Rev. Mod. Phys. \textbf{81}, 109 (2009).
\bibitem{Sim97} S. H. Simon and P. A. Lee, Phys. Rev. Lett. \textbf{78}, 1548 (1997).
\end{thebibliography}
\end{document}